Title: Clinical Validation of a Real-Time Machine Learning-based System for the Detection of Acute Myeloid Leukemia by Flow Cytometry

Authors:

Lauren M. Zuromski[1,*], Jacob Durtschi[1,*], Aimal Aziz[2], Jeffrey Chumley[2], Mark Dewey[1], Paul English[1], Muir Morrison[1], Keith Simmon[3,4], Blaine Whipple[2], Brendan O'Fallon[1,4], David P. Ng[1,2,4]

*Co-first authors

[1]Department of Applied Artificial Intelligence and Bioinformatics, Institute for Research and Innovation, ARUP Laboratories, Salt Lake City, Utah, USA

[2]Hematologic Flow Cytometry, ARUP Laboratories, Salt Lake City, Utah, USA

[3]Institute for Clinical and Experimental Pathology, ARUP Laboratories, Salt Lake City, Utah, USA

[4]Department of Pathology, University of Utah, Salt Lake City, Utah, USA

Abstract

Machine-learning (ML) models in flow cytometry have the potential to reduce error rates, increase reproducibility, and boost the efficiency of clinical labs. While numerous ML models for flow cytometry data have been proposed, few studies have described the clinical deployment of such models. Realizing the potential gains of ML models in clinical labs requires not only an accurate model, but infrastructure for automated inference, error detection, analytics and monitoring, and structured data extraction. Here, we describe an ML model for detection of Acute Myeloid Leukemia (AML), along with the infrastructure supporting clinical implementation. Our infrastructure leverages the resilience and scalability of the cloud for model inference, a Kubernetes-based workflow system that provides model reproducibility and resource management, and a system for extracting structured diagnoses from full-text reports. We also describe our model monitoring and visualization platform, an essential element for ensuring continued model accuracy. Finally, we present a post-deployment analysis of impacts on turn-around time and compare production accuracy to the original validation statistics.

## Introduction

The present state of laboratory medicine is characterized by declining reimbursements coupled with a decreasing workforce and increasing costs. To combat these trends, numerous avenues have been pursued, including automation and consolidation of laboratories. However, the past decade has seen an explosion of interest in the use of machine learning (ML) in medicine. Numerous examples of ML applications in clinical medicine, radiology, and pathology have been published. Unfortunately, few examples have seen the utilization in the field. There are many reasons for this including a lack of infrastructure and skilled personnel in this new field. Indeed, in clinical flow cytometry, utilization of these tools in the laboratory is in its infancy at best and, to our knowledge, there has not been a published instance of a clinically deployed AI model for use in the flow cytometry laboratory.

Within the flow cytometry literature, several algorithms have been developed and proposed(Ng et al., 2015; Zhao et al., 2020; Ng & Zuromski, 2021; Mallesh et al., 2021; Simonson et al., 2021; Salama et al., 2022; Bazinet et al., 2023; Lewis et al., 2024; Lu et al., 2024; Shopsowitz et al., 2024), each with advantages and disadvantages in computational complexity, explainability, and training. Within our high-volume national reference laboratory, the confluence of clinical use cases, conservative culture, and technical feasibility provide distinct advantages in developing an ML pipeline that could predict the presence of Acute Myeloid

Leukemia in initial triage tubes and inform technologists of the need to perform additional studies to fully characterize this disease. While Simonson et al.(Simonson et al., 2022) described a mock-up of such a system for use in triggering CLL add-ons in a reference laboratory setting, their study was only able to predict cases in real time without any effect in the clinical workflow as their project was stymied by the lack of a laboratory information system (LIS) and user interface that could alert technologists to perform testing. Likewise, Lu et al. (Lu et al., 2024) demonstrated their clinical validation for cell population enumeration on a small number of cases in autoimmune lymphoproliferative syndrome, however their system was not clinically implemented due to issues with LIS integration (private communication with authors). Other studies have demonstrated good performance in the detection, and in some cases subclassification, of acute myeloid leukemia but similarly did not implement their systems for clinical use (Zhong et al., 2022; Monaghan et al., 2022; Lewis et al., 2024).

In brief, our clinical workflow consists of a panel of three 10-color screening tubes (Table 1) run on all peripheral blood and bone marrow samples. Like most flow cytometry laboratories, interpretation of this screening panel is done manually by technologists and pathologists and a decision is made to perform further testing to better characterize an abnormal cell population. In the case of acute myeloid leukemia (AML), if >15% myeloid blasts are seen in a bone marrow or peripheral smear (considering the new World Health Organization (WHO)(Khoury et al., 2022) and International Consensus Classification (ICC)(Arber et al., 2022) classifications in addition to the well described discordance between flow and morphologic blast enumeration(Chen et al., 2022)), then additional panels are performed to assess lineage assignment, erythroid and megakaryocytic differentiation, and other blasts aberrancies.  Within our lab, this process takes an average of 27 hours due to delays in having a technologist review the triage panels, followed by performing the required add-on studies.  Additionally, within our homegrown LIS, such cases are flagged as expedited with special notification on the work queue for analysts and pathologists. Through this work we sought to address these inefficiencies in our flow cytometry workflow through the development and validation of an ML pipeline to automate the screening step in triaging AML positive specimens for further analysis.  Importantly, we demonstrate the clinical implementation of this pipeline and provide empirical insights to the impact of our tool at production level.

## Methods

Our implementation consists of four primary components. Most important is the machine learning model itself, which takes the flow cytometry event data as inputs and produces a prediction for AML.  Supporting the machine learning model is the computational infrastructure that triggers the model under appropriate conditions, manages and transforms data, and stores results.  An additional component aggregates past data and constructs visualizations that are reviewed by laboratory staff and medical directors to detect model drift or other potential failures. Finally, because training the model requires categorical labels associated with each sample, we train an additional large language model to parse raw report text and identify samples that are AML positive. Each of these four components is described in detail below.

Instruments and Data

Our lab operates five 10-color NaviosEX (Beckman Coulter Miami, FL) cytometers, which produce approximately 250,000 viable events based on forward and side scatter characteristics per tube. Cytometer outputs, initially in modified FCS file format (*.LMD format), are transformed into comma separated values (CSV) and an accompanying metadata file, both of which are uploaded to cloud storage (see section X or figure Y for details). A single sample consists of event data from three tubes, which we refer to as the 'B-cell', 'T-cell', and 'Myeloid' tubes (Table 1). Data analyzed in this paper were clinical samples generated between 18 Feb 2021 and 1 May 2024 with a Train/Test/Validation period between 18 Feb 2021 and 31 Dec 2022.

| Panel | FITC | PE | ECD | PeCy5.5 | PeCy7 | APC | APC-A700 | APC-A750 | PB | KO |
|---|---|---|---|---|---|---|---|---|---|---|
| B cell | Kappa | Lambda | CD19 | CD5 | CD23 | CD10 | CD38 | CD200 | CD20 | CD45 |
| T cell | CD57 | CD7 | CD4 | CD5 | CD2 | CD56 | CD8 | CD3 | CD16 | CD45 |
| Myeloid | CD14 | CD33 | CD13 | CD117 | CD11b | CD38 | CD34 | CD64 | HLA-DR | CD45 |

Table 1: Markers analyzed per tube

Model Overview

Our machine learning model is composed of two elements: a self-supervised feature extraction and aggregation module that learns a low-dimensional representation of the tube data, and a supervised classification module trained to detect AML. The first module is implemented using Self-Organizing Maps (SOMs) (Van Gassen et al., 2015; Zhao et al., 2020; Mallet et al., 2021). For each event, the SOM projects the 13 parameter values to two dimensions corresponding to positions on a 32x32 array. Events for a single tube are aggregated via mean pooling to produce a single 32x32 array representing the entire tube. Separate SOMs are trained for each of the B-cell, T-cell, and Myeloid tubes. The classification module consists of a gradient-boosted decision tree that consumes the flattened and concatenated SOMs and produces the final AML prediction result. The classifier is trained with labels extracted from raw report text (see section 'Model details: Classifier training' ).

Model details: SOM training

SOM training consists of unsupervised learning of event embeddings from high-dimensional space onto a 2D grid. The trained SOM models can then be used to project new datasets with matching spatial mappings. Our SOM training data set consisted of 13,566 samples from February 2021 to May 2022, and SOMs were trained for 30 epochs. The result of SOM training is a 2D centroid grid, in our case, 32x32. A separate SOM was trained for the B-cell, T-cell, and myeloid tube data.

Each tube data set consisted of approximately 250,000 events. Training data was chosen from a contiguous date range that did not include any test or validation samples to assure data independence with the test data set. SOM training data was restricted to samples without any clinically significant findings as this appeared to improve overall predictive performance. The SOM model code was based on python QuickSOM package(Mallet et al., 2021; Bouvier, 2024), with JAX implementation to provide GPU-acceleration for faster processing(Bradbury et al., 2024).

Model details: Classifier training

The supervised classifier is a gradient boosted decision tree as implemented in the python XGBoost package(Chen & Guestrin, 2016). SOM features from all three tubes are flattened and concatenated to produce a single vector of length 3072. The feature sets for all training samples were used with four-fold, train-test splits with three repeats in a round of hyperparameter optimization of the XGBoost python classifier model using the Hyperopt python package (Bergstra et al., 2013). The chosen hyperparameters were then used in a round of feature selection based on most-commonly-used features among all 12 train-test split repeats. Specifically, features with a feature importance gain greater than zero in at least 8 of the 12 repeats of the training process were identified and used in the final model training. This reduced the feature set from 3072 per sample to 396. Feature selection was followed by another round of hyperparameter tuning. The feature selection process, hyperparameter tuning, and final training were optimized for the $F_1$ statistic. The full training data set of SOM projections was then used to train the final XGBoost model with the chosen hyperparameters and feature subset.

For classifier training, samples were divided into temporally distinct train, test, and validation sets (Table 2). We chose a temporal split strategy, with all training data preceding all validation samples, to better emulate the conditions under which the algorithm is intended to operate in production. The test sample set was used to establish the threshold for final determination of the AML positive class, while the validation set was used to calculate recall (also known as sensitivity), precision (positive predictive value), and related statistical measures at the threshold established using the test set.

| Sample set | Date Range | Number of Samples | AML-positive samples (% of samples) |
| --- | --- | --- | --- |
| Training | Feb-18-2021 – May-29-2022 | 13,566 | 470 (3.46%) |
| Test | June-01-2022 – July-31-2022 | 1,349 | 49 (3.6%) |
| Validation | Aug-01-2022 – Dec-31-2022 | 3,464 | 113 (3.26%) |

Table 2: Classifier training dataset overview.

Model details: Inference

Flow cytometer event data for each sample's B-cell, T-cell, and myeloid tubes were each embedded via the respective SOM model. All events are embedded independently and merged with average pooling to create a single SOM representation for each tube.

A subset of the values from each representation are then used as input features in the AML model as described in the AML classifier training section above. The AML classifier inference then provided a classifier score between zero and one indicating confidence that the sample is AML positive. Samples with scores above a clinically and operationally determined threshold were labeled AML positive (0.9 in our model and dataset).

Training Label Generation

Patient reports in our lab are stored as free text. To generate discrete, categorical labels suitable for training the classifier we used a two-step process. First, a corpus of approximately 6,000 reports was manually annotated by a board-certified pathologist (DPN). Next, a support vector machine (SVM)(Joachims, 1998) was trained on the corpus of reports to generate preliminary diagnosis labels for unreviewed chart text. For support vector classification, the text data was first converted into a matrix of token counts that is then transformed into a term frequency-inverse document frequency (TF-IDF) representation(Salton & Buckley, 1988), which measures the importance of tokens, or words in the reports, emphasizing unique words and reducing the impact of words that occur frequently. The SVM achieved an $F_1$ score of 0.90 for AML across a 5-fold cross-validation.

Although the SVM model achieved high performance, our team has since implemented a higher performing model based on the BERT architecture[25], which we use to produce labels for model monitoring purposes (see section Monitoring Model Performance). Specifically, we use the BiomedBERT (Chakraborty et al., 2020) variant which has been fine-tuned on biomedical literature. This model was trained on ~25k reports (including manually reviewed reports and reports classified as high confidence by the TF-IDF/SVM model) to generate labels for the validation set. Pre-trained BiomedBERT model weights were obtained from HuggingFace, a collaborative hub for sharing deep learning models and data. The BiomedBERT model was then fine-tuned in a supervised setting on our labeled report data. The fine-tuned BiomedBERT model achieved an $F_1$ score of 0.93 on 1836 testing cases. These preliminary, model-generated labels are reviewed by residents, fellows, and/or Medical Directors in an in-house, and internally hosted, web application to confirm or correct the preliminary model label.

Clinical Compute and Data Storage Infrastructure

Because our model is used to support clinical operations, we require infrastructure that is resistant to hardware failures and network disruptions, can scale up and down appropriately with sample volume, and that tracks system events, software versions, and related sample metadata. To meet these requirements, we developed a cloud-based, containerized architecture with a minimal on-premises footprint (Figure 1). Raw instrument data is written to a local volume as List Mode Data (LMD) files, and a software daemon monitors the filesystem for new files. The LMD files are preprocessed to remove protected health information (PHI) and uploaded to Amazon Web Services' Simple Storage Service (AWS S3). After each LMD file is uploaded, a message is sent via AWS Simple Queue Service (SQS) to an AWS Lambda function.  The lambda function checks if all three LMD files – one for each tube type - for a given sample have been uploaded, and if so, triggers the machine learning workflow. Workflows are managed by Argo, a Kubernetes-native workflow manager that allows for easy tracking, visualization, and scheduling of multistep workflows. The Argo workflow transfers the cytometry files from object storage to a compute node, executes the model, and stores the results. Model predictions and associated data (model version, workflow inputs, container id, etc.) are stored in a document database and exposed via a GraphQL API.  A final message is published to AWS Simple Notification Service (SNS) triggering downstream applications – which typically subscribe via another SQS queue – to gather the data and display it to users.

The use of AWS Lambda and Argo to manage these workflows allows full elasticity, scaling to zero between runs and scaling up as necessary when many LMD files land concurrently. SQS and Argo allow for automatic retries and resumption of tasks that fail due to temporary infrastructure outages and rolling updates of the service. The API that exposes the data also contains utility endpoints for manually re-triggering the workflow in case of a long-lasting infrastructure outage, or for development and debugging purposes.

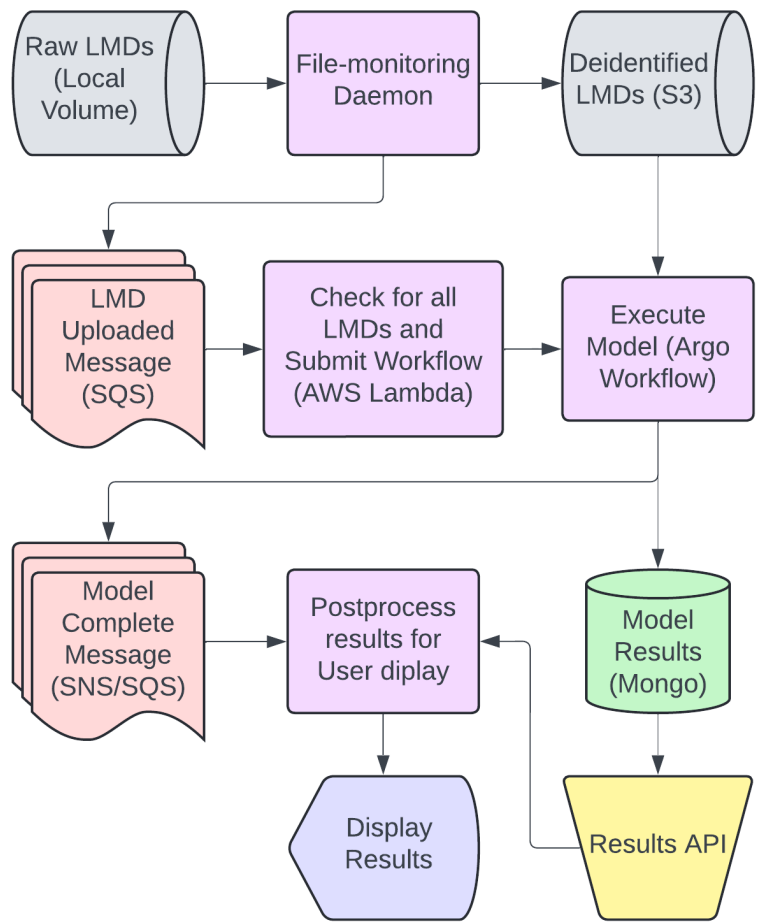

Figure 1: Operational software infrastructure and workflow for flow cytometry samples.

Monitoring Model Performance

Clinical application of machine learning requires tools to monitor model performance over time. To facilitate easy review of important metrics we constructed a system to aggregate and visualize performance statistics. A Tableau dashboard was created to monitor model performance (Figure 2). Sample labels are obtained from a MongoDB collection storing both the BiomedBERT predictions and the manually reviewed sample labels, where the reviewed labels are used when available and the BiomedBERT predictions serve as a secondary source to complete the corpus of labels for comparison with the model predictions. The flow model prediction (using a positive threshold of 0.9) is compared to these true labels and the number of true positives, false positives, true negatives, and false negatives are computed and displayed. These prediction classifications and the predictions are also shown in a table per accession. The chart text from the reports are shown on hover over rows in the table; if a subject matter expert (SME), in this case a board-certified hematopathologist, notices a discrepancy between the chart text and the label, the SME can correct the label in our in-house web application (as described in above) and the changes will take effect in the dashboard within 12 hours (the source data is updated twice daily). Weekly model metrics ($F_1$, sensitivity, specificity, negative predictive value, and positive predictive value) are displayed as line graphs to monitor any model performance trends. These same metrics are listed in a table for the overall model performance. Daily average positive prediction values are displayed as a line graph with the $95^{th}$ and $5^{th}$ percentiles of overall positive predictions shown as horizontal lines to help visualize if positive predictions drift over time akin to Levey-Jennings control plots.

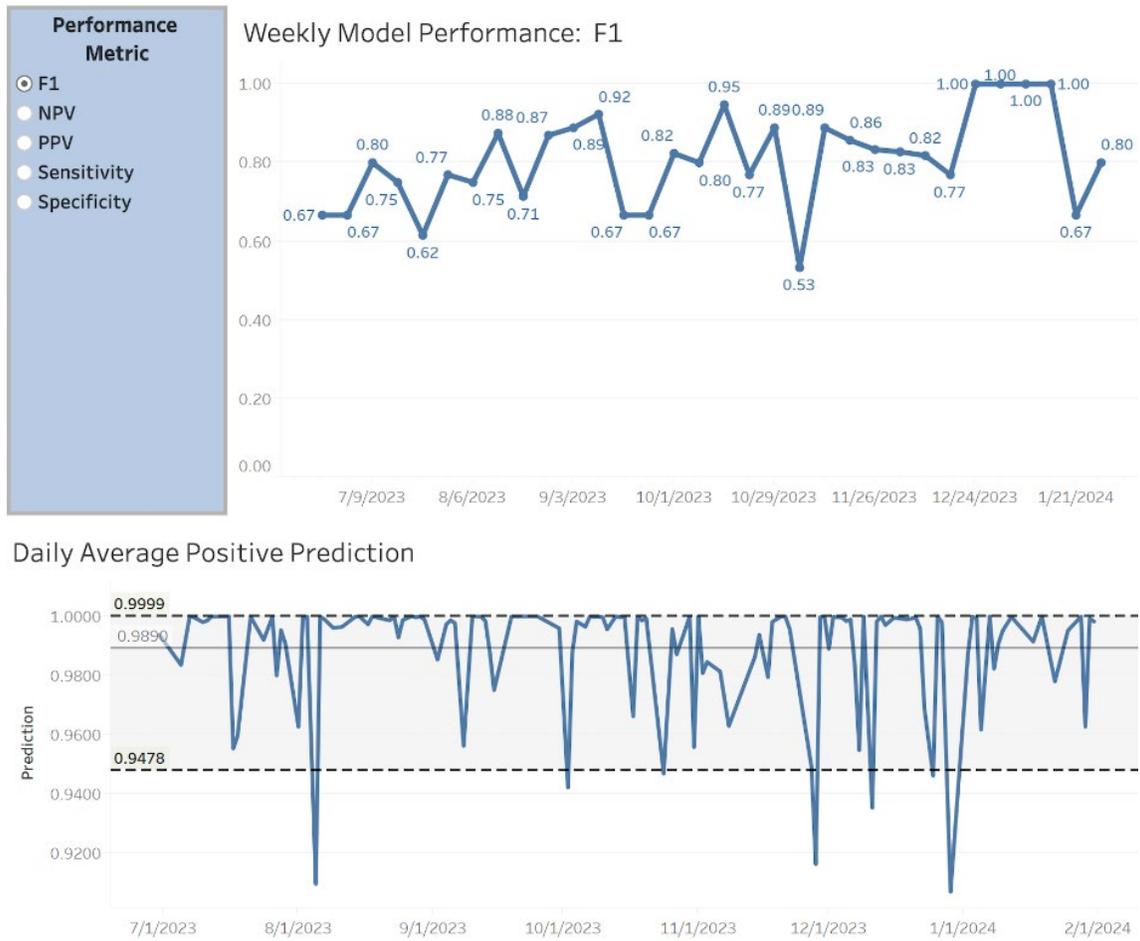

Figure 2: Weekly model performance and daily average positive prediction values over time, as displayed in our model-monitoring dashboard.

## Results

Sensitivity and Precision

Using the test sample set, we constructed and manually reviewed a precision-recall curve for AML detection accuracy. A threshold of 0.9 was chosen based on clinical

Private Information

and operational factors to favor precision and specificity, which yielded 40 true positives, 1 false positive, 9 false negatives and 1298 true negatives in the test set.

Validation set samples were classified with the developed model and the chosen classifier model threshold (0.9) to generate the overall metrics of sensitivity of 76.99%, precision 92.55%, and specificity 99.79%. However, several samples were of low quality such as hemodilute, paucicellular, suboptimal, or clotted. Accuracy metrics were generated after removal of these samples (Table 3). ROC plots and classifier score histograms of the trimmed datasets are shown in Figures 3 and 4. After poor-quality samples were removed sensitivity increased to 80.77% at precision of 92.31%, while specificity remained unchanged at 99.79%.

| Dataset | Full validation dataset | After poor-quality samples removed |
| --- | --- | --- |
| Sensitivity | 0.7699 | 0.8077 |
| Precision | 0.9255 | 0.9231 |
| Specificity | 0.9979 | 0.9979 |

Table 3: Sensitivity, Precision and Specificity before and after removal of low-quality samples including hemodilute, paucicellular or clotted.

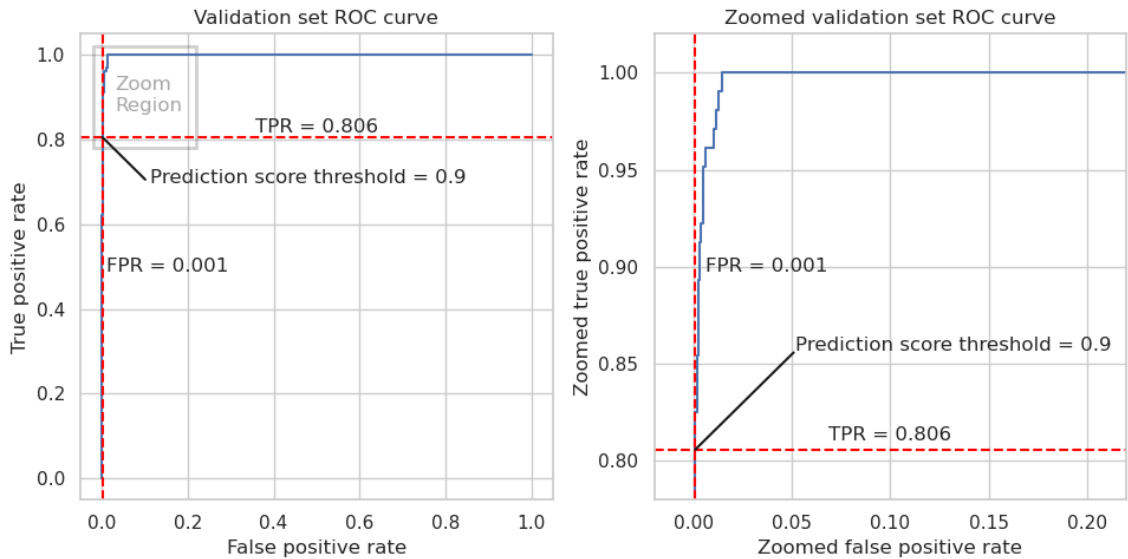

Figure 3: Validation set ROC curve with false positive rate (FPR) and true positive rate (TPR) for chosen prediction score threshold of 0.9 marked. Full range ROC shown on left and zoomed upper left corner of ROC on right.

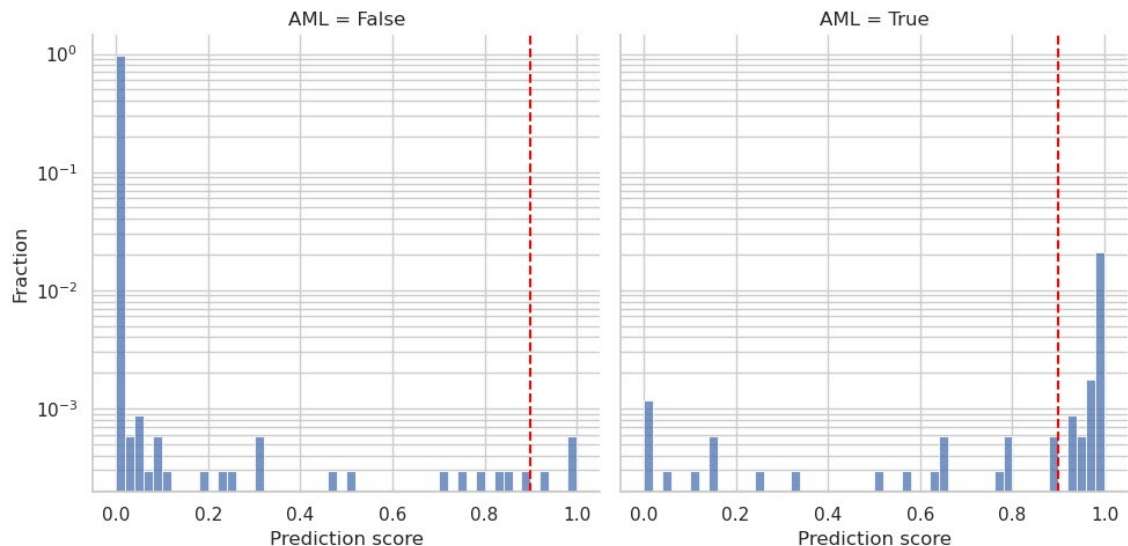

Figure 4: Validation AML positive/negative prediction score distributions for validation samples with chosen prediction score threshold of 0.9 marked. Note that the y axis is in logarithmic scale and that most samples are low prediction score, AML negative samples in the bin to the far left.

Reproducibility

To compute within-instrument reproducibility, eight samples were each run in triplicate on a single flow cytometer. For between-instrument reproducibility, the eight samples were each run on either three or four different flow cytometers. Coefficients of variation (CV) of the model prediction scores were calculated separately for the between-instrument and within-instrument repeats. The eight reproducibility samples included four known AML negatives and four known AML positives. Figure 5 shows individual sample repeat prediction scores. Note that the False Negative (FN) AML sample (AML 2) has prediction scores much higher than other negative samples, indicating the model identified a difference in this specimen but not with sufficient confidence to classify it as AML positive. One AML known negative was manually diagnosed as near borderline AML (Low AML). Prediction scores for this sample were near zero but slightly elevated compared to other AML negative samples.

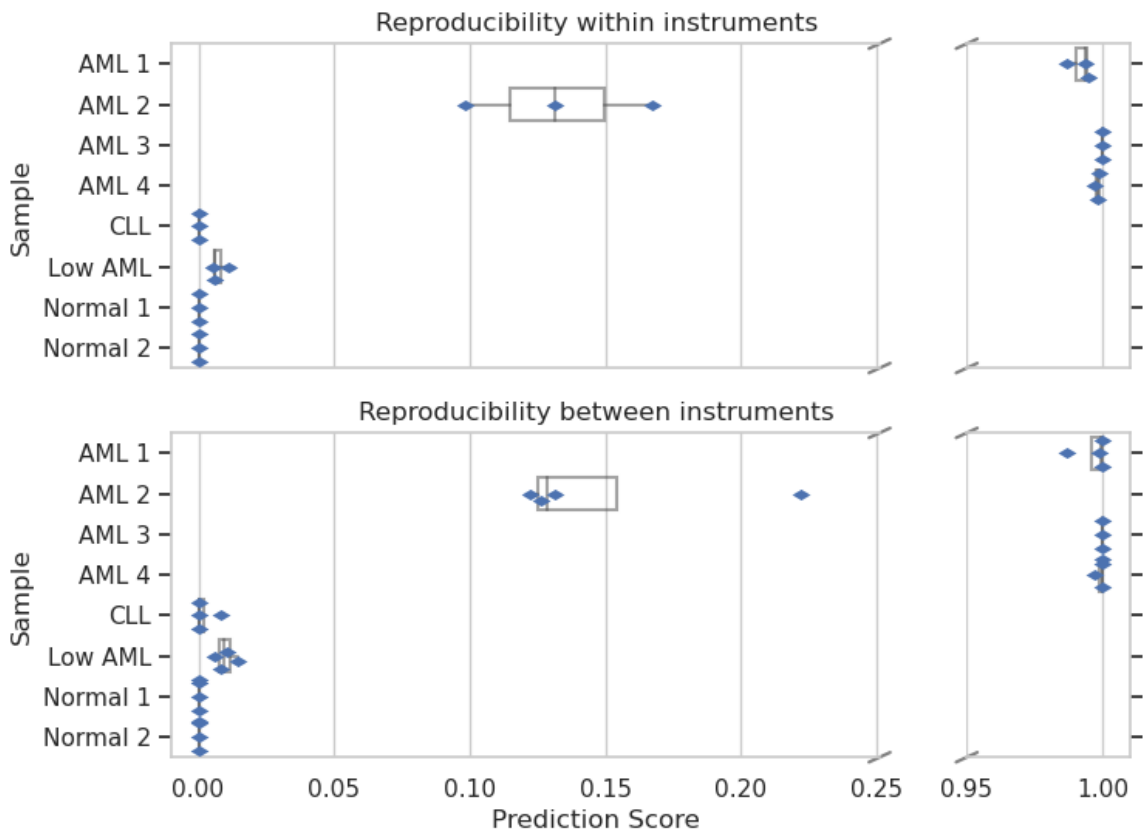

Figure 5: Within and between-instrument AML prediction score reproducibility for 8 samples with a range of ground truth diagnoses. The X axis is broken as all scores were near zero or one. One false negative AML sample was included, AML 2, which had consistently higher scores than the true negative AML samples in these plots but still well below the chosen threshold of 0.9.

Performance on Demographic Subgroups

From samples processed between July 1, 2023 and April 30, 2024, the proportion of correct calls to overall calls, or accuracy, was computed and then compared in a pairwise manner between subgroups of age (10-year bins from 30 to 80 years of age, while binning 0 to 29 and 80+ years of age for low decadal sample sizes), sex, and self-reported race using two-proportion z-tests (Figure 6). Resulting p-values were adjusted for multiple comparisons using the Benjamini-Hochberg method to control the false discovery rate. Though the model accuracy for 50- to 59-year-olds at 99.5% was statistically significantly higher (p = 0.01) than for 80+ year olds at 98.2%, this difference of 1.3% (95% CI: 0.6% to 2.0%) is not practically significant. All demographic subgroups had accuracies >=95%, indicating no hidden demographic model bias.

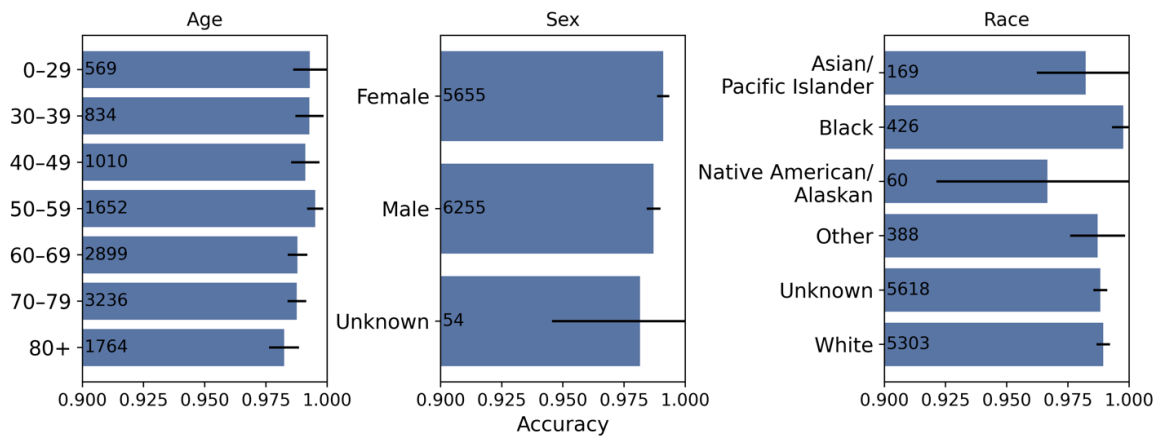

Figure 6: Accuracy by age, sex, and race with 95% confidence interval bars. Subgroup sample sizes are annotated on the bars.

Retrospective review of performance metrics in production

| Dataset | Full validation dataset | Real-world use, June 2023-April 2024 |
| --- | --- | --- |
| Sensitivity | 0.7699 | 0.69 +/- 0.03 |
| Precision | 0.9255 | 0.939 +/- 0.018 |
| Specificity | 0.9979 | 0.9989 +/- 0.0003 |

Table 4: Performance metrics of our AML classifier on our validation dataset and in actual clinical use. Uncertainties +/- one standard deviation, as estimated by non-parametric bootstrap.

Taking the predictions of our AML classifier from mid-June 2023 to the end of April 2024, we computed sensitivity, precision, and specificity as reported in Table 4. We find that precision and specificity are essentially identical with the results from our validation set, within uncertainty, while real-world sensitivity is approximately 8% lower (69% vs 77% computed in validation). To see if there were detectable trends in these metrics over time that might be suggestive of model drift, we binned our samples by calendar month and computed these metrics over the same time period, as shown in Figure 7. Overall, we observe no clear trends and suspect the fluctuations are simply due to small-number statistics; as there are < 50 AML cases in a typical month, the difference between 1 and 2 false positives is a substantial difference in precision. (Note the complete absence of false positives in June, August, September, and December 2023, leading to no bootstrap estimation of uncertainties for precision and specificity.)

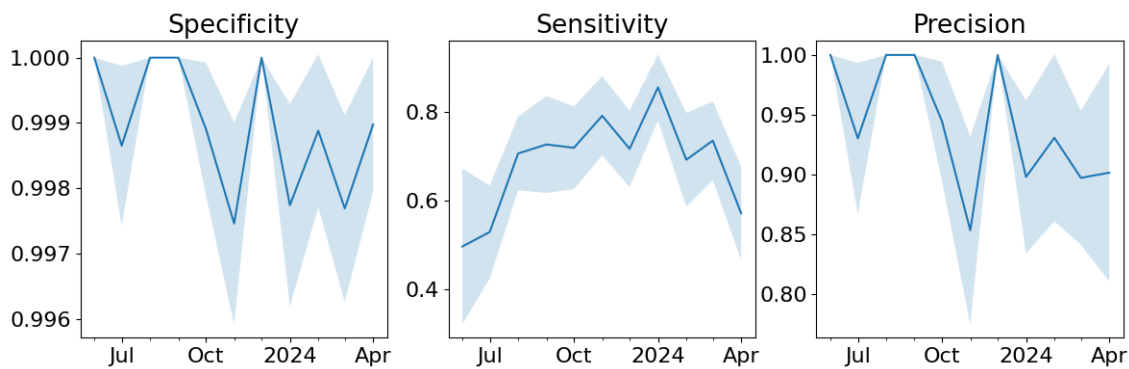

Figure 7: Specificity, sensitivity, and precision of our AML classifier in actual use, binned by month. Uncertainty bands are +/- 1 standard deviation, estimated by non-parametric bootstrap.

Effects of assay drift on classifier results

As noted, the system showed good performance through April 2024, however a noticeable and transient decline in $F_1$ scores was observed in early May of 2024 which temporally correlated with laser failures on two instruments. To determine the relationship between these results, we determined the channel means of all markers and found a significant drop in CD45 Krome-Orange was correlated with the changes in $F_1$ performance (Figure 8). On the contrary, other markers such as CD33, CD34, and CD117 also showed drops in channel means, however they are not correlated with changes in $F_1$ performance suggesting that the expression levels of CD45 is a major contributor to overall classifier performance.

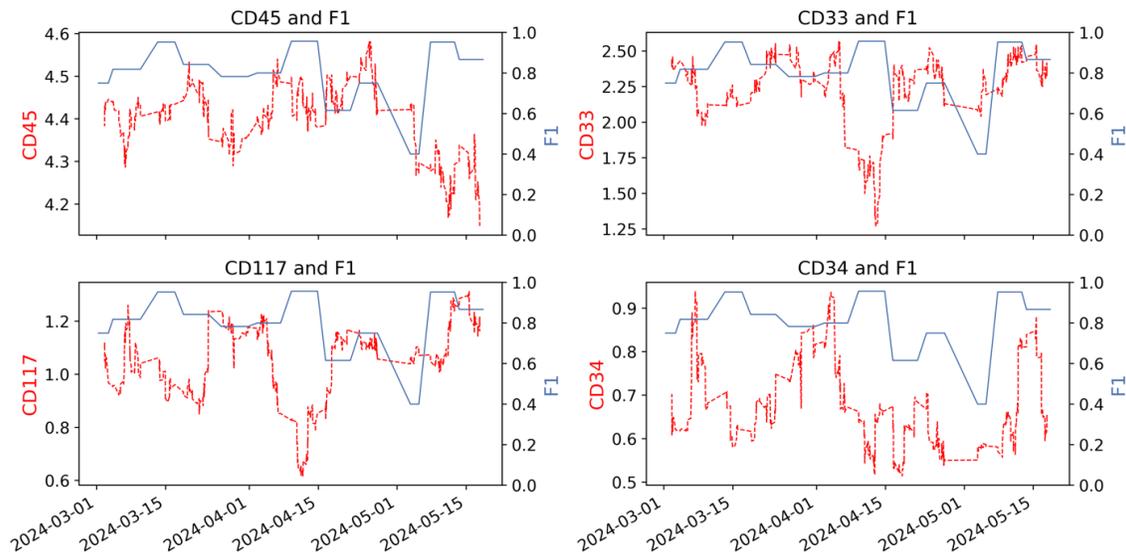

Figure 8: Time series plots of weekly $F_1$ values (blue solid lines) with rolling averages of CD45, CD33, CD117, and CD34 signals (red dashed lines).

Turn Around Time

The in-laboratory turnaround time (TAT) of AML and missed-AML cases (using our labels as ground truth) were extracted from 12 June 2023 (i.e. the date the model was put into production) though 1 May 2024. A 30-day moving average filter was applied to these TATs and the resulting difference in weekly average AML and missed-AML TAT were then computed. To control for variations in laboratory TAT due to various external causes (e.g. staffing, weather, instrumentation availability) and to isolate the potential time savings of predicting AMLs, the difference in weekly average TAT between missed-AML and AML cases was computed. This difference had no significant trends over the period of interest ($p > 0.05$).

Retrospective review of False Positive Cases

FCS data from false positive cases between 5 July 2023 to 20 March 2024 were manually reviewed along with their reports by a board-certified pathologist (DPN). Out of the 8753 bone marrow and peripheral blood cases analyzed by the pipeline, 13 apparent false positive (FP) cases were identified. Four FP cases (0.046% of all cases) were misannotated by the BiomedBERT text classifier (i.e. the reports and data were those of corresponding to AMLs). Six FP cases (0.069%) demonstrated between 8 and 14% myeloid blasts upon manual review of the text and primary flow data wherein our lab standard for triggering add-on studies is 15% myeloid blasts. Two FP cases (0.022%) showed significant plasma cell populations (71% and 15% respectively), which interestingly showed coexpression of CD56 and CD117 without CD19 and diminished CD45 sitting within the CD45/SS defined 'blast' gate. One FP case (0.011%) was a clear B-lymphoblastic leukemia/lymphoma with coexpression of CD34 with monocyte intensity CD33 and the usual B cell level of CD19 without CD10. An analysis of the 96 false negative cases showed there were a combination of cases with low viability cases or near our 15% blast cutoff with the highest blast count being 40%.

## Discussion

In this work, we describe the development, validation, and implementation of a bespoke solution for the identification of critical leukemias using a novel combination of LLM text extraction and self-supervised embedding steps that provide inputs for the ultimate supervised classification of AML in flow cytometry specimens. We observed clinically acceptable analytical performance in validation, with precision of 92.55% and recall of 76.99%. Furthermore, while many studies have looked at retrospective performance analogous to our study during clinical validation, we were also able to evaluate prospective/operational performance over a 10-month period.  While precision and specificity remained high without significant variation, sensitivity was slightly worse with a 6% decrease compared with retrospective validation.  While this stability was promising, two older instruments suffered laser issues that resulted in an overall decrease in CD45 signal in May 2024.  These instruments were taken out of service and repaired; however, the downward trend is visible even in samples that were in control per the manufacturer specifications.  This decreased CD45 signal was strongly correlated with a drop in the $F_1$ score, suggesting that assay/cytometer drift could have a significant effect on classifier performance.  We are investigating alternative classifiers based on point cloud or transformer networks(Lewis et al., 2024) which may be more robust to assay drift.  Data normalization methods in flow cytometry have been described(Hahne et al., 2010; Van Gassen et al., 2020), however as these are batch normalization techniques, their utility for real-time data normalization may be limited.

After implementation, we were unable to show a significant decrease in overall TAT. Several confounding process improvement initiatives likely contributed to the lack of impact, including the introduction of an all-digital workflow system on 2 Jan 2024, a priority queue for critical cases on 29 MAR 2024, and autoverification (removal of a technologist step between pathologist sign-out and resulting in the LIS) on 4 APR 2024.  Within a specific metric (i.e. time to availability for pathologist review), we observed a trend downwards over highly variable data.  We will continue to accrue data and look prospectively for natural experiments that may help us determine the effect of this process.

Bias in clinical AI is a critical concern and an object of active research.  Most papers we reviewed did not report on the effect of age, sex and race on classifier performance.  We were able to determine that no significant changes were seen between demographic groups with some caveats.  Age and sex are common demographic identifiers and trivial to extract from our clinical dataset, however, self-identified race was difficult to assess for a variety of well-described reasons(Fang et al., 2019; Johfre et al., 2021), and for a large swath of the dataset (46.96%, 5618/11964), it was simply unavailable. Therefore, an unobserved bias is possible. Indeed, the obtainable racial makeup of our dataset is skewed towards our home state, which has a higher proportion of white patients than the rest of the United States(Anon, n.d.). Nevertheless, given the size and diversity of our reference laboratory clients (no single institution comprises more than 20% of our dataset), our training data's demographic makeup is likely similar to our prospective data described here and we anticipate it more closely resembles that of the United States.

It is well understood that running the same file through a non-stochastic process such as most computer algorithms should result in the same results. However, the collection of flow cytometry events is indeed a stochastic process drawing a sample of events from some distribution. Even as the number of events collected increases, there are still subtle variations in the data driven by a myriad of causes, including specimen set up conditions, and electronic and detector noise. Under these conditions, we examined the effects of repeated sample runs through our wet bench process on the pipeline's predictions and observed good consistency. This supports the assertion that wet bench variation does not affect classifier performance to any significant degree.

Finally, the regulatory landscape is fuzzy at best. The recent promulgation of the FDA final rule on laboratory developed tests (LDT) included a grandfather clause under which this test modification resides(Anon, 2024). Unfortunately, future developments are likely stymied requiring either 501k submissions, or submission of this test modification to the New York Department of Health (NY CLEP). While the FDA has issued guidance documents (Bogdanoski et al., 2024) which suggest that all AI tools must seek FDA clearance, the categorization of these tools remains unclear and their pathway to clinical implementation will require reassessment as regulations become more defined.

To our knowledge, this is the first description of a clinically deployed ML algorithm in the flow cytometry laboratory. From proof of concept, this project took over two years to reach deployment: including development, validation, and building of additional user and data interfaces necessary to support the pipeline. System architecture in a complex reference laboratory setting with numerous legacy systems is a non-trivial task and other smaller laboratories with less complex informatics environments may be able to implement these systems faster. The true clinical impact on TAT is unclear and will require more time to accrue data. Finally, our data and pipeline are geared only towards predicting on our lab developed triaging panels and the generalizability of our ML model is likely minimal. Further work on developing generalized models including those that can assimilate add-on studies as well as predicting on other useful diagnostic categories is underway and shows great promise.

Private Information

Acknowledgements

We thank Dr. Alexandra Rangel for her assistance in editing this manuscript for publication and the entire hematologic flow cytometry laboratory for its assistance in bringing this pipeline online.

Conflict of Interest Statement

The authors declare no conflict of interest.